\documentclass[aps,pre,twocolumn,groupedaddress,reprint,showpacs]{revtex4-1}

\usepackage[]{graphicx}
\usepackage{amsmath}

\usepackage{hyperref}
\usepackage{url}

\newcommand{\wvalue}{w_2}
\newcommand{\mscore}{q_m}
\newcommand{\mscoresingle}{q_{m,m^\prime}}

\begin{document}

\title{Motif-based success scores in coauthorship networks are highly sensitive
  to author name disambiguation}
\thanks{published as \href{http://dx.doi.org/10.1103/PhysRevE.90.032811}{Phys.~Rev.~E {\bf90}, 032811 (2014)}}

\author{David F. Klosik}
\email[]{klosik@itp.uni-bremen.de}

\author{Stefan Bornholdt}
\email[]{bornholdt@itp.uni-bremen.de}
\affiliation{Institute for Theoretical Physics, University of Bremen,
  Hochschulring 18, 28359 Bremen, Germany}
\author{Marc-Thorsten H\"utt}
\email[]{m.huett@jacobs-university.de}
\affiliation{School of Engineering and Science, Jacobs University Bremen,
  Campus Ring 1, 28759 Bremen, Germany}

\begin{abstract}
Following the work of Krumov~\textit{et al.}~%
\href{http://dx.doi.org/10.1140/epjb/e2011-10746-5}{%
[Eur.~Phys.~J.~B \textbf{84},~535 (2011)]} %
we revisit the question whether the usage of large citation datasets
allows for the quantitative assessment of social (by means of coauthorship of
publications) influence on the progression of science. Applying a more
comprehensive and well-curated dataset containing the publications in the
journals of the American Physical Society during the whole 20th century we find
that the measure chosen in the original study, a score based on small induced
subgraphs, has to be used with caution, since the obtained results are highly
sensitive to the exact implementation of the author disambiguation task.
\end{abstract}


\maketitle

\section{Introduction}

Ever since the seminal work of Kuhn \cite{Kuhn} it is widely accepted that the
institutional process of knowledge production, i.e.~science, cannot be fully
described in purely logical, content-related terms, but has a significant social
aspect to it.  However, although the scientific community provides a
comprehensive bookkeeping of its efforts by citing earlier work in new
publications, only recently has this information been made widely accessible in
the form of electronic datasets. With the aggregated citation information within
a set of scientific publications at hand, one might now be able to
quantitatively assess the extent to which the social embedding of science
influences its structure and progression. %

Traditionally, the focus of citation data analysis has been on the single
publication level; indeed, the most prominent property of \textit{paper citation
  networks} in which the vertices represent papers while directed edges
represent the citations between the publications has been described by de Solla
Price as early as 1965 \cite{deSollaPrice1965}: The number of citations a paper
receives (i.e.,~the corresponding node's in-degree) is broadly distributed,
rendering highly cited publications significantly more frequent than they would
be if scientists cited earlier work randomly. Similar broad degree distributions
have been found in networks describing, e.g.,~technical, social, or biological
interactions \cite{BarabasiAlbertReview2002}, and the question which might be
the mutual underlying process, led to a now well-studied model class for network
growth governed by a \textit{rich-get-richer} effect, commonly referred to as
preferential attachment
\cite{deSollaPrice1976,BarabasiAlbert1999,DorogovtsevMendesSamukhin_2000_PRICE,%
  DorogovtsevMendes_2001_continuous,DorogovtsevMendes_2000_aging,
  BianconiBarabasi_2001,KrapivskyRedner_2001,KrapivskyRednerLeyvraz_2000}.
In more recent times, citation data itself has mainly been used to quantify the
assessment of scientific research thereby interpreting citations as indicators
of impact or assignment of credit. To this end, numerous quantitative measures
have been described which range from counting direct citations to considering
also indirect citation paths \cite{RadicchiFortunatoCastellano2008,
  Bollen_etal_2009, chen_etal_2006, walker_etal_2007, klosik_citation_2013} and
some of which aim at the scholar \cite{Hirsch2005,Radicchi_et_al_2009} or the
journal level \cite{West_etal_2010}.  Additionally, there have also been
structural investigations, e.g.,~regarding the community level
\cite{chen_redner_2010} or other topological properties such as the richness of
feed-forward loops in citation networks \cite{WuHolme_2009}.

Another line of research has considered the collaboration network that can be
constructed from citation data given the authorship metadata. More precisely, in
a \textit{coauthorship network} vertices represent authors and are connected by
an undirected link if the two corresponding authors have coauthored one or more
papers together. These networks have been investigated about a decade ago
\cite{newman_collaboration_I,newman_collaboration_II} with the main findings
being the rather broad degree distribution and the strong small-world effect,
i.e., the short paths between scholars in the network. %
There are also few approaches combining both citation and collaboration data,
e.g.,~\cite{martin_coauthorship_2013}. %

In this study we will construct a coauthorship network from a citation dataset,
while the actual citation information is used to assess the success of the
resulting links in the network. %

\begin{figure}[h]
  \centering
  \includegraphics[width=1.\columnwidth]{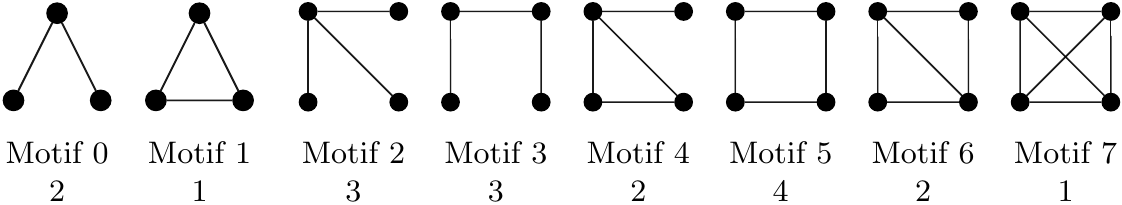}
  \caption{The subgraphs used in this study. In the bottom line the minimum
    number of distinct papers required to build the subgraph is given.}
  \label{fig:motifskizze}
\end{figure}

While there have been studies about large-scale properties of coauthorship
networks such as path lengths or community sizes and, of course, on the
single-vertex scale, following Krumov \textit{et al.} \cite{krumov_motifs_2011}
here the focus will be on the intermediate level of small induced subgraphs
known from the investigation of network motifs which originated in the
biological context \cite{milo_motifs_2002, milo_superfamilies_2004} and whose
application has drawn some considerable criticism since the null-model graph
ensemble has to be chosen with great care
\cite{fretter_subgraph_2012,beber_artefacts_2012}.  Note, however,
that throughout this study \textit{score} does not refer to the
otherwise commonly used \textit{z score}, since we are not interested
in subgraph frequencies, but to the $\mscore$ value that is defined
below.  We will focus on the three- and four-node undirected induced
subgraphs shown in Fig.~\ref{fig:motifskizze}.

We construct the collaboration network from a citation dataset provided by the
American Physical Society (\textit{see Materials and Methods}) by, first,
identifying the articles' authors given in the article metadata. All author
instances (i.e.,~names in author lists) are then grouped so that, ideally, all
instances corresponding to the same actual scholar are represented by a single
vertex in the network. %
There are different possible implementations of this \textit{author
  disambiguation task} which has become an object of investigation in its own
right \cite{schulz_exploiting_2014,silva_network-based_2013} with methods
ranging from directly comparing the author names to incorporating metadata or
even citation data. %
While in Krumov \textit{et al.}~one specific version has been used, here we find
that the score proposed to assess the correlation between the collaboration
pattern and the success of the corresponding papers is very sensitive to the
exact implementation of the author disambiguation. %
We stress that this kind of ambiguity is not restricted to coauthorship networks
but is also present in other network applications (see Sec.~\ref{chap:concl}).
In the bibliography context, however, some effects of disambiguation errors have
been pointed out in \cite{fegleyTorvik_2013}.

Furthermore, we show how the score is affected by the exclusion of large
collaborations when the length of the author list above which papers are
discarded is varied; in \cite{krumov_motifs_2011} this value was fixed at a
number of eight.

\section{Computation} 
\begin{figure}[]
  \centering
  \includegraphics[width=1.\columnwidth]{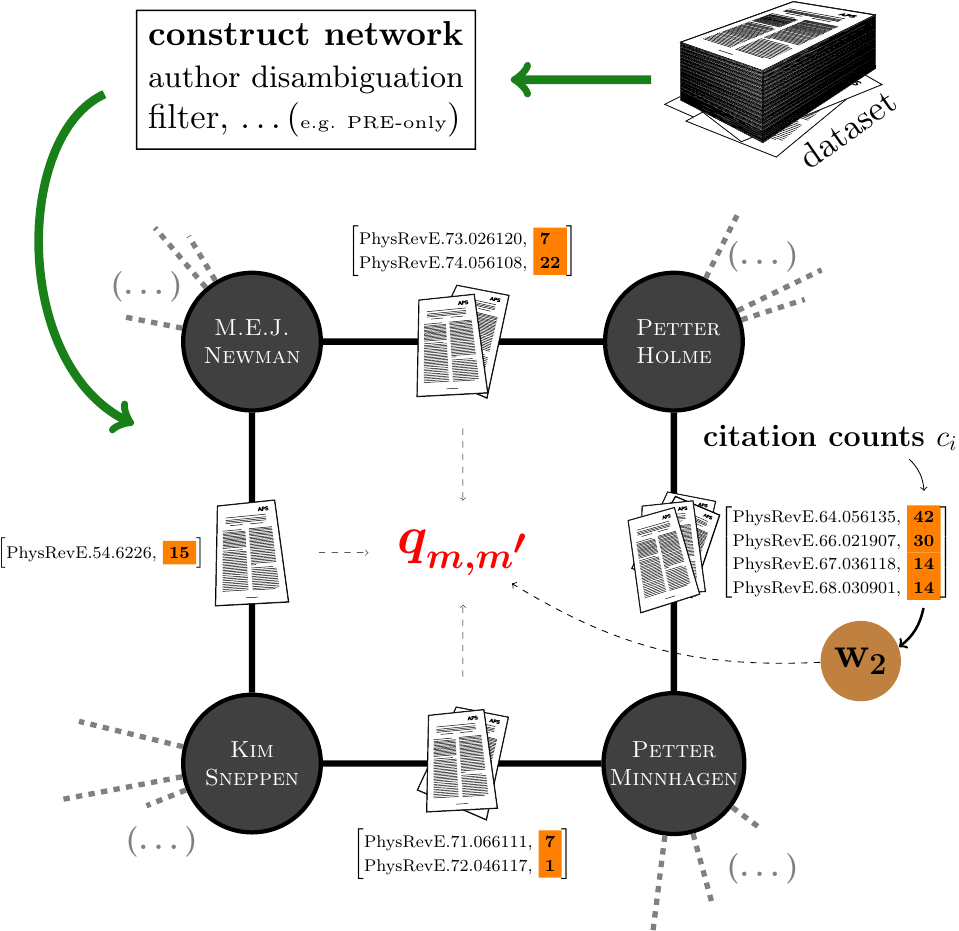}
  \caption{(Color online) A sketch of the score computation.}
  \label{fig:skizze}
\end{figure}

First, we review the computation of the score proposed by Krumov \textit{et al.}
Two distinct authors who have coauthored are connected by an edge, $e$, that
represents the list of all their mutually published papers, $P(e)$. %
Note that by this procedure a single publication can be represented by many
edges.  In order to assess whether there is a correlation between the
collaboration structure on the small subgraph scale 
and the scientific impact of the publications contained in the subgraph's edges
the number of citations to those papers is used to, first, compute the average
number of citations to the papers of a single edge
\begin{equation}
  \wvalue = \langle c \rangle_e =  \frac{1}{|P(e)|} \sum_{p \in P(e)} c(p)
  \label{eq:edge_score}
\end{equation}
and then take the average over all edges of all instances of a specific motif, 
$M_m$, i.e.,~
\begin{equation}
  \mscore = \frac{1}{N_m} \sum_{m^\prime \in M_m} q_{m,m^\prime} = \\
  \frac{1}{N_m E_m} \sum_{m^\prime \in M_m} \sum_{e \in m^\prime} \\ 
  \langle c \rangle_e
\end{equation}
where $N_m=|M_m|$ gives the number of instances of the subgraph of type $m$ and
$E_m$ yields the number of its edges (e.g.,~for the box motif $E_5=4$). %
A sketch of the procedure is given in Fig.~\ref{fig:skizze}.  If one now
shuffles the citation numbers among all publications the $\mscore$ is uniformly
distributed, indicating that the motif-scale collaboration patterns in the given
coauthorship network and the papers' success are not correlated after
reshuffling.  With the particular choice of ($\ref{eq:edge_score}$) the motif
scores are expected to yield the average number of citations of all papers in
the network for every motif. %
The fact that this holds in all our computations implies that we obtain proper
averages of the often broadly distributed citation numbers.  In
\cite{krumov_motifs_2011} alternative edge scores have been proposed but with
the current dataset ($\ref{eq:edge_score}$) turns out to be the appropriate
choice.  It is important to notice that the shuffling procedure only affects the
citation data on the edges of the otherwise fixed collaboration network,
i.e.,~no topological shuffling is applied.

Krumov \textit{et al.}~report considerably higher $\mscore$ scores for the
four-node motif called Motif $5$ in Fig.~\ref{fig:motifskizze} which they call
the \textit{box motif}. This subgraph stands out since it needs four distinct
publications to be constructed (while e.g.~the four-node clique, Motif $7$,
might contain only one single mutually published paper) and there must not be
any other collaborations between the four authors than in the author pairs
corresponding to the four edges. %
The box motif therefore is an anticlustered structure and considerably fewer
instances are found than of the other motif types. Again, we stress that we do
not focus on motif frequency.

In order to keep the network topology and the dynamical quantity separated, in
contrast to \cite{krumov_motifs_2011} we do not discard edges which are composed
only of papers that did not receive any citation, but keep them in the network.

\subsection{Maximum collaboration size} 
As mentioned above, in \cite{krumov_motifs_2011} all papers with more than eight
authors have been discarded.  Here we allow for different values of the
\textit{maximum collaboration size} (MCS) and investigate its influence on the
$\mscore$ score (Sec.~\ref{chap:mcs}). %
In terms of the social aspects encoded in a coauthorship network the exclusion
of very large collaborations can be argued for since the thousands of links
between their authors will hardly represent the same degree of personal
connection as an edge between, e.g.,~two scholars publishing in a team of two.

\subsection{Author name disambiguation}
While one of the more elaborate author disambiguation schemes might have been
used here, in order to assess the sensitivity of the score proposed in
\cite{krumov_motifs_2011} we instead chose two versions of a very simple measure
which consider only the authors' names as provided in the author lists of the
publications in the dataset \cite{Milojevic2013}.  In the \textit{all initials}
(allINIT) disambiguation two author instance names are considered identical if
in addition to the surname all initials are compatible (meaning that
\textsc{J.~Smith} and \textsc{John Smith} are merged). %
The second implementation requires the full first name strings to match; we
consider this \textit{strict} (STRICT) since it will separate author instances
if the first name is abbreviated in one and written out in the other case like
the ones from the above example. %
However, there are surnames common enough to be shared by people with different
given names; if these given names have compatible initials the allINIT method
will spuriously merge them. In order to address this we apply a third
disambiguation which we will refer to as SPLIT: We apply independently the
allINIT and STRICT disambiguation to the same data and then track those
allINIT names that are split into more than two distinct authors in the STRICT
implementation. All papers with those names among the authors are then
filtered. On the resulting dataset we then perform all remaining steps of the
calculation with the allINIT method applied.
Note that the result of this procedure depends on all previous data filtering,
especially the one according to the MCS, so the
number of remaining papers has to be compared to the number of publications in
the original data after discarding those with more than MCS authors. In the
cases shown here the data had been prepared with a MCS value of $10$ before the
application of the SPLIT filtering and we find that while for the merge of the
journals %
$36\%$ and in PRC $32\%$ are filtered, in the remaining single journal portions
around $80\%$ of the papers are kept. 
For comparison, if one excluded all papers from the dataset that had at least
one author whose given name is only provided in initialized form not only would
over half of the dataset be excluded, but also this filtering would be biased
against large collaborations since there name abbreviation is used to save
printing space.

\section{Materials and Methods}
The citation data used in this study is composed of all APS journals published
between July 1893 and December 2009 and the associated information about
citations between these papers, and may therefore be assumed to cover many major
20th century contributions to physics. The dataset can be requested at the
American Physical Society \footnote{See
  \url{https://publish.aps.org/datasets}.}.  The data is provided in separate
files corresponding to the individual APS journals which emphasize different
topics; we chose to exclude \textit{Reviews of Modern Physics} and the
\textit{online-only} journals \textit{Physical Review Special Topics -
  Accelerators and Beams} and \textit{Physics Education Research} due to their
considerably smaller size compared to the remaining journals.
These remaining journals vary in size from \textit{Physical Review E} with
$35022$ to \textit{Physical Review B} with $133269$ publications.  In addition
to extensive bibliographic metadata such as authorship, publication history, and
PACS numbers the dataset also provides \textit{article type} tags, thereby
allowing us to filter nonstandard material (i.e.,~those tagged \textit{comment,
  erratum, reply, editorial, essay, publisher-note, retraction, miscellaneous})
and restrict our analysis to standard research publications.

Since motif enumeration is computationally costly, for most of the calculations
the RANDESU algorithm as described by Wernicke \cite{wernicke} is applied which
instead of enumerating all subgraphs samples uniformly from the set of all
motifs. By performing duplicate runs of the motif score computation with the
same parameters we checked that the scores are not sensitive to the sampling
procedure. For smaller networks the full enumeration is performed by
application of the ESU algorithm, presented also in \cite{wernicke}.

\section{Results}
We computed the $\mscore$ score in both the single journal portions and the
whole APS dataset and found that the shuffling of the citation frequencies among
the papers indeed yields a uniform distribution of the scores. As mentioned
above, in all cases the appropriate edge weight to achieve this was $w_2=\langle
c \rangle_e$, i.e.,~the averaged citation frequency of all papers on the
corresponding edge.  In the following we show the influence of a variation of
the maximal number of authors above which a paper is excluded from the data, as
well as the sensitivity of the $\mscore$ score against different implementations
of the author disambiguation task.

\subsection{MCS scan}
\label{chap:mcs}
\begin{figure}[h]
  \includegraphics[width=1.\columnwidth]{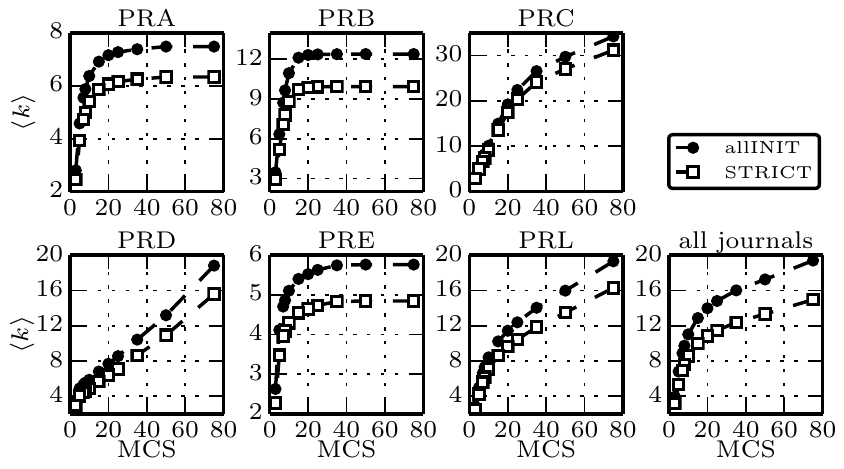}
  \caption{A scan of the MCS value translates to a variation of the average
    degree.}
  \label{fig:mcsscan}
\end{figure}

In Fig.~\ref{fig:mcsscan} the average network degree as a function of the MCS
value is shown for both the allINIT and the STRICT disambiguation. %
Increasing the MCS value translates into the introduction of new potential nodes
and edges to the network, and since a publication with $a$ authors can produce
up to $a(a-1)/2$ new edges $\langle k \rangle$ grows. %
Unlike PRA, PRB and PRE, the journals PRC, PRD and PRL which feature larger
collaborations do not show a saturated average degree in the shown MCS
interval. The fact that the STRICT disambiguation results in more distinct
authors than the allINIT implementation is reflected by the lower average
degrees.

\begin{figure}[h]
  \includegraphics[width=1.\columnwidth]{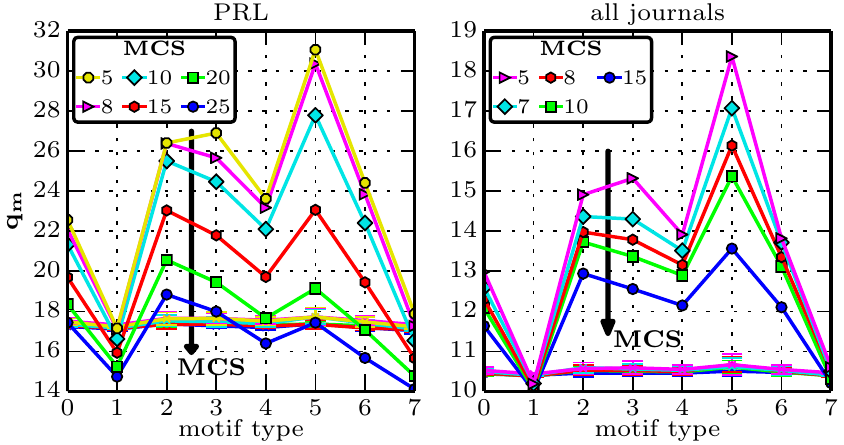}
  \caption{(Color online) Two example MCS scans (STRICT disambiguation). The box
    motif signal is weaker for larger MCS. %
The almost flat lines correspond to the averaged results of $30$ runs with
shuffled citation counts.}
  \label{fig:mcsscanscores}
\end{figure}

Especially with the STRICT disambiguation scheme we can reproduce the main
finding of \cite{krumov_motifs_2011}, the higher $\mscore$ score for the box
motif, for small enough values of the maximum collaboration size. %
Although in few journal portions this result is rather robust, in most cases the
box motif signal tends to decrease with increasing MCS;
Fig.~\ref{fig:mcsscanscores} shows two examples: %
In the merge of the journals the box motif keeps the highest score in the given
MCS range, while in PRL the box motif signal not only gets less pronounced but
is lost at $\text{MCS}\approx 15$.

\subsection{Influence of the author disambiguation}
\begin{figure}[h]
  \centering
    \includegraphics[width=1.\columnwidth]{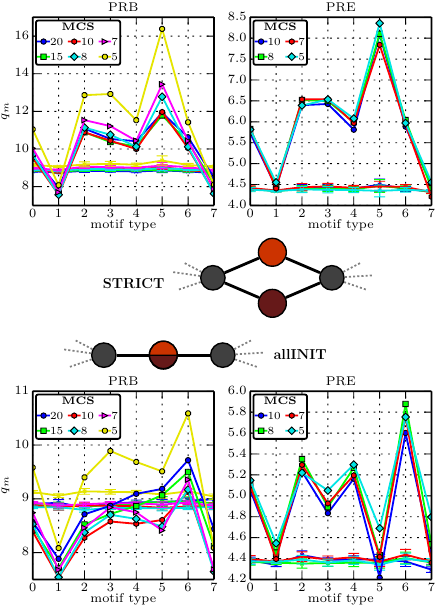}
  \caption{(Color online) The influence of the local topological perturbation
    introduced by switching between the STRICT (top) and the allINIT (bottom)
    disambiguation scheme for the PRB and the PRE single journal
    portions. Again, the almost flat lines correspond to the averaged results of
    $30$ runs with shuffled citation counts.}
  \label{fig:disambcomposite}
\end{figure}

The exact implementation of the author disambiguation turned out to be crucial
considering the $\mscore$ distribution. From the network perspective it can be
interpreted as a local perturbation (as illustrated in the sketch in
Fig.~\ref{fig:disambcomposite}) which strongly influences the $\mscore$ scores
computed on the few-node subgraph scale.
While the SPLIT disambiguation shows a rather similar behavior to the STRICT
case (only PRA and PRL do not show a maximum score for the box motif in the
former case), by switching from the STRICT to the allINIT disambiguation scheme
the $q_m$ score distribution can change qualitatively. For example, the journal
merge does not show the box motif signal any longer and also in the case of PRB
and PRE shown in Fig.~\ref{fig:disambcomposite} the box motif scores are
suppressed.

\subsection{Distributions of the $\mscoresingle$ scores}
\begin{figure}[h]
  \centering
  \includegraphics[width=1.\columnwidth]{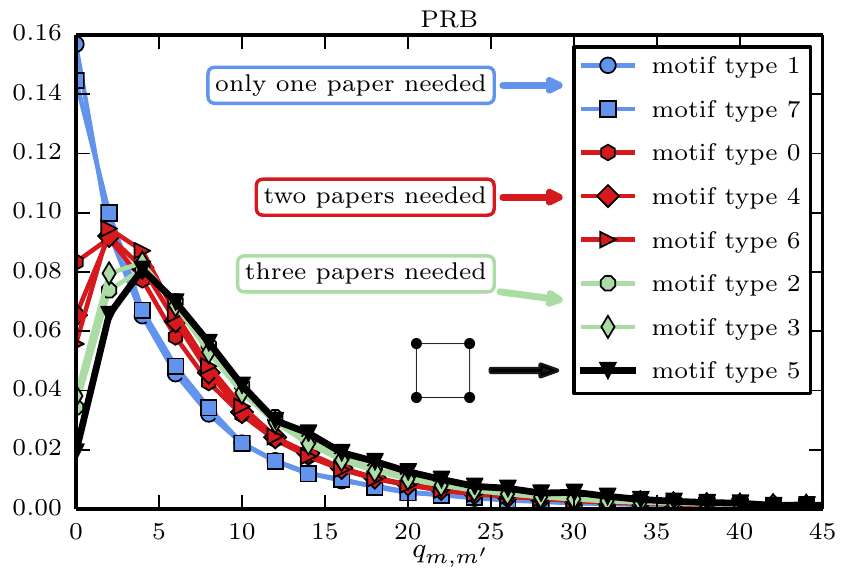}
  \caption{(Color online) Normalized frequency of the $\mscoresingle$ scores for
    the different motif types. The PRB journal portion has been used with
    $\text{MCS}=10$ and the SPLIT filtering has been applied.}
  \label{fig:distributionPRB}
\end{figure}

The $\mscore$ are averaged values over the $\mscoresingle$ scores of the single
instances of the specific motif type $m$ and as such are influenced of course
not only by the top-ranked but also by the motif instances with lowest scores.
In Fig.~\ref{fig:distributionPRB} the distributions of the $\mscoresingle$
values for the different motif types are shown for the example of the PRB
network portion.  In the depicted range the distributions can be grouped
according to the minimal number of distinct papers required to construct the
specific motif (bottom row in Fig.~\ref{fig:motifskizze}).
In the box motif which requires the largest minimum number of distinct papers
very low $\mscoresingle$ scores are suppressed while the three- and four-node
cliques which need only one mutual publication among the authors often show very
small scores.  This is due to the distribution of the citation counts used in
the computation of the scores: The many poorly cited publications can translate
directly to low scores for cliques while it is unlikely for the at least four
edges of a box motif to exclusively consist of poorly cited papers.  Indeed, the
$\mscoresingle$ distributions of the motif types $0,4,6$ and $2,3$ which require
the intermediate number of $2$ and $3$ edges, respectively, is consistent with
this interpretation.
This renders the investigation of the top motif instances according to the
$\mscoresingle$ scores less promising.

A paper with $a$ authors contributes to $a(a-1)/2$ edges which in turn may
contribute to very many motif instances. One corresponding effect can be seen
when examining the top motif instances according to the $\mscoresingle$ scores:
These lists can be dominated by highly-cited publications. For example, more
than half of the top motifs in the PRA journal portion (SPLIT disambiguation,
$MCS=10$) share the publication \textit{``Quantum computation with quantum
dots''} by DiVincenzo and Loss \cite{PhysRevA.57.120}.
If one is not, however, interested in compiling lists of top instances, but only
in averaged values like the $\mscore$, another approach might be used.  The
$\mscore$ for a specific motif type $m$ can be rephrased as a weighted sum over
the citation counts of all papers that contribute to instances of that motif;
the weights, however, depend not only on how often the paper contributes but
also on the numbers of other papers it shares its edges with:
\begin{align}
  q_m =& \sum_{i \in P} c_i  \; \frac{1}{N_m E_m} \sum_{m^\prime \in M_m} 
  \sum_{e \in m^\prime} \frac{1}{|P^{(i)}(e)|}\, , \\ 
  &|P^{(i)}(e)|=\infty \quad \mbox{ if } i \notin P(e) \notag
\end{align}
with $P$ being the set of all papers.  %
A simple alternative would be to set these weights to $\delta_{i\in M_m}/|P_m|$
where $P_m$ is the set of all papers that at least once contribute to a subgraph
instance of type $m$, i.e.,~to compute the average citation frequency of the
papers that contribute to the motif type in question but consider each
publication only once.

\begin{figure}[h]
  \centering
  \includegraphics[width=\columnwidth]{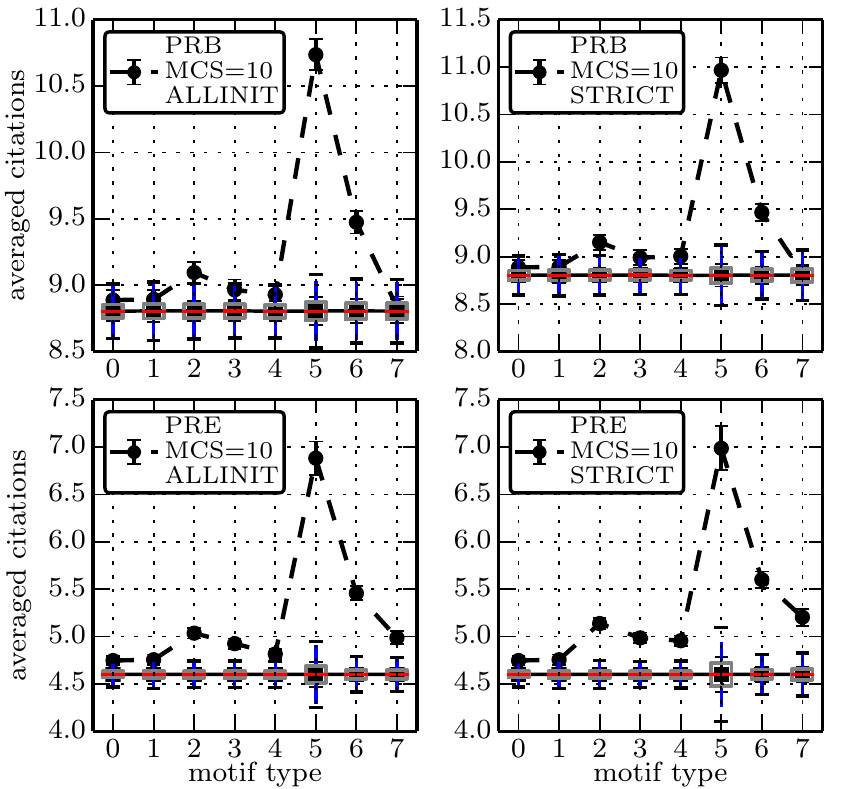}
  \caption{(Color online) The average citation counts of the papers that
    contribute to at least one instance of a given motif type for the example of
    the PRB and PRE data.  Compare to Fig.~\ref{fig:disambcomposite} where also
    the allINIT and the STRICT disambiguation is shown. }
  \label{fig:alternativeScore}
\end{figure}

Indeed, the box motif shows the highest average citation counts compared to the
other motif types for all single journals as well as the merge of all journals
independently of the choice of the disambiguation scheme and for MCS values not
smaller than $5$.
In Fig.~\ref{fig:alternativeScore} these citation averages are given for the
example of the PRB and the PRE data and an MCS value of $10$ (compare to
Fig.~\ref{fig:disambcomposite}).  In the figure a box plot of the average of a
random sample of size $|M_m|$ drawn from the citation distribution is given,
showing that the average citation count for the box motif lies significantly
above the sample average and is therefore not just due to the smaller sample
size.
However, since uniform sampling assumes that every paper is a part of every
motif type with the same probability, the sampling procedure might be refined to
incorporate the specific conditions imposed by the different motif types.  In a
first estimation, we restricted the samples for the box motif to those papers
that have at least two coauthors with at least degree $2$ each (i.e.,~have at
least one other collaborator) for the case of the SPLIT disambiguation.
Although this increases the sample average, the measured average citation count
is still several (at least $5$) standard deviations above the sample average.

\section{\label{chap:concl}Discussion and Conclusion}
Motivated by the question to what extent small-scale patterns of collaboration
among scholars influence the recognition of the resulting papers within the
scientific community (in terms of the number of citations the publications
receive) we searched induced few-node subgraphs (known from motif analysis) in a
comprehensive citation dataset provided by the American Physical Society.  We
then assigned a score to the detected subgraphs as described in Krumov
\textit{et al.}~\cite{krumov_motifs_2011} based on the citation counts of the
corresponding publications so that an average value, $\mscore$, for every
subgraph type $m$ can be computed. %
It turns out that this score is sensitive to certain details of the network
aggregation from the data and the main result of the aforementioned study
\cite{krumov_motifs_2011}, i.e.,~the highest score for the anticlustered
box motif, cannot be seen clearly in all cases: For example, discarding
different amounts of the largest collaborations which translates to a variation
of the average network degree influences the $\mscore$ signature.  A stronger
effect can be seen when the network is locally perturbed by changing the
implementation of the author disambiguation task; in the \textit{all initial}
method of author disambiguation the box-motif score can be suppressed, while a
stricter disambiguation scheme yields a high score. Consequently, such a score
should be used with care.  This result might be of interest in other
applications of few-node induced subgraph-based scores to data which is
incomplete or not unambiguously processed into a graph representation since it
highlights a possible difficulty on the motif scale in addition to the
well-known problems concerning the proper null-model selection in the usual
motif-analysis.
While the ambiguity concerning whether two author names are to be translated
into one or two nodes is exclusive to coauthorship networks, several other
types of complex networks have related underlying disambiguation tasks:
Enzyme-centric metabolic networks (where the nodes are enzymes and a link
represents common substrates and products) can also be formulated on the level
of reactions (see, e.g., \cite{beber_artefacts_2012}); for many organisms, the
enzyme inventory of the genome is incomplete
\cite{Feist:2009ik,Schellenberger:2010kb} and the enzyme-to-reaction mapping
introduces similar issues as the author disambiguation. On the basis of the
present investigation we expect that the topological properties of
enzyme-centric and reaction-centric metabolic networks show similar differences
as the coauthorship networks under disambiguation schemes of different
stringency.
For bacteria, transcriptional regulatory networks (where the nodes are genes and
a directed link from gene $g1$ to gene $g2$ represents the regulation of a $g2$
by a transcription factor produced by $g1$) are sometimes more adequately
described on the level of groups of genes, called \textit{operons}, under common
regulation (see, e.g., \cite{Marr:2008p4284}). In eukaryotic organisms, attempts
to apply this concept lead to ambiguities \cite{blumenthal2004}, similar to
those encountered in the author disambiguation.
In computational neuroscience, the topology of cortical area networks is often
analyzed on the basis of ``regions of interest'' (ROI) from diffusion spectrum
imaging, rather than functionally defined cortical areas
\cite{Hagmann2008}. Relating these ROI to the functional cortical areas, and
comparing results obtained with different segmentations of the cerebral cortex,
again, is comparable to the author disambiguation task.

\begin{acknowledgments}
We acknowledge the support of the Deutsche Forschungsgemeinschaft
(DFG) under Contracts No.~BO 1242/6-1 and No.~HU 937/9-1.
\end{acknowledgments}

\bibliographystyle{apsrev4-1}
\bibliography{./references}

\begin{thebibliography}{41}%
\makeatletter
\providecommand \@ifxundefined [1]{%
 \@ifx{#1\undefined}
}%
\providecommand \@ifnum [1]{%
 \ifnum #1\expandafter \@firstoftwo
 \else \expandafter \@secondoftwo
 \fi
}%
\providecommand \@ifx [1]{%
 \ifx #1\expandafter \@firstoftwo
 \else \expandafter \@secondoftwo
 \fi
}%
\providecommand \natexlab [1]{#1}%
\providecommand \enquote  [1]{``#1''}%
\providecommand \bibnamefont  [1]{#1}%
\providecommand \bibfnamefont [1]{#1}%
\providecommand \citenamefont [1]{#1}%
\providecommand \href@noop [0]{\@secondoftwo}%
\providecommand \href [0]{\begingroup \@sanitize@url \@href}%
\providecommand \@href[1]{\@@startlink{#1}\@@href}%
\providecommand \@@href[1]{\endgroup#1\@@endlink}%
\providecommand \@sanitize@url [0]{\catcode `\\12\catcode `\$12\catcode
  `\&12\catcode `\#12\catcode `\^12\catcode `\_12\catcode `\%12\relax}%
\providecommand \@@startlink[1]{}%
\providecommand \@@endlink[0]{}%
\providecommand \url  [0]{\begingroup\@sanitize@url \@url }%
\providecommand \@url [1]{\endgroup\@href {#1}{\urlprefix }}%
\providecommand \urlprefix  [0]{URL }%
\providecommand \Eprint [0]{\href }%
\providecommand \doibase [0]{http://dx.doi.org/}%
\providecommand \selectlanguage [0]{\@gobble}%
\providecommand \bibinfo  [0]{\@secondoftwo}%
\providecommand \bibfield  [0]{\@secondoftwo}%
\providecommand \translation [1]{[#1]}%
\providecommand \BibitemOpen [0]{}%
\providecommand \bibitemStop [0]{}%
\providecommand \bibitemNoStop [0]{.\EOS\space}%
\providecommand \EOS [0]{\spacefactor3000\relax}%
\providecommand \BibitemShut  [1]{\csname bibitem#1\endcsname}%
\let\auto@bib@innerbib\@empty
\bibitem [{\citenamefont {Kuhn}(1996)}]{Kuhn}%
  \BibitemOpen
  \bibfield  {author} {\bibinfo {author} {\bibfnamefont {T.~S.}\ \bibnamefont
  {Kuhn}},\ }\href@noop {} {\emph {\bibinfo {title} {The Structure of
  Scientific Revolutions}}}\ (\bibinfo  {publisher} {University of Chicago
  Press, Chicago},\ \bibinfo {year} {1996})\BibitemShut {NoStop}%
\bibitem [{\citenamefont {Price}(1965)}]{deSollaPrice1965}%
  \BibitemOpen
  \bibfield  {author} {\bibinfo {author} {\bibfnamefont {D.~J. D.~S.}\
  \bibnamefont {Price}},\ }\href {\doibase 10.1126/science.149.3683.510}
  {\bibfield  {journal} {\bibinfo  {journal} {Science}\ }\textbf {\bibinfo
  {volume} {149}},\ \bibinfo {pages} {510} (\bibinfo {year}
  {1965})}\BibitemShut {NoStop}%
\bibitem [{\citenamefont {Albert}\ and\ \citenamefont
  {Barab\'asi}(2002)}]{BarabasiAlbertReview2002}%
  \BibitemOpen
  \bibfield  {author} {\bibinfo {author} {\bibfnamefont {R.}~\bibnamefont
  {Albert}}\ and\ \bibinfo {author} {\bibfnamefont {A.-L.}\ \bibnamefont
  {Barab\'asi}},\ }\href {\doibase 10.1103/RevModPhys.74.47} {\bibfield
  {journal} {\bibinfo  {journal} {Rev. Mod. Phys.}\ }\textbf {\bibinfo {volume}
  {74}},\ \bibinfo {pages} {47} (\bibinfo {year} {2002})}\BibitemShut {NoStop}%
\bibitem [{\citenamefont {Price}(1976)}]{deSollaPrice1976}%
  \BibitemOpen
  \bibfield  {author} {\bibinfo {author} {\bibfnamefont {D.~J. D.~S.}\
  \bibnamefont {Price}},\ }\href {\doibase 10.1002/asi.4630270505} {\bibfield
  {journal} {\bibinfo  {journal} {J. Amer. Soc. Inform. Sci.}\ }\textbf
  {\bibinfo {volume} {27}},\ \bibinfo {pages} {292} (\bibinfo {year}
  {1976})}\BibitemShut {NoStop}%
\bibitem [{\citenamefont {Barab\'asi}\ and\ \citenamefont
  {Albert}(1999)}]{BarabasiAlbert1999}%
  \BibitemOpen
  \bibfield  {author} {\bibinfo {author} {\bibfnamefont {A.-L.}\ \bibnamefont
  {Barab\'asi}}\ and\ \bibinfo {author} {\bibfnamefont {R.}~\bibnamefont
  {Albert}},\ }\href {\doibase 10.1126/science.286.5439.509} {\bibfield
  {journal} {\bibinfo  {journal} {Science}\ }\textbf {\bibinfo {volume}
  {286}},\ \bibinfo {pages} {509} (\bibinfo {year} {1999})}\BibitemShut
  {NoStop}%
\bibitem [{\citenamefont {Dorogovtsev}\ \emph {et~al.}(2000)\citenamefont
  {Dorogovtsev}, \citenamefont {Mendes},\ and\ \citenamefont
  {Samukhin}}]{DorogovtsevMendesSamukhin_2000_PRICE}%
  \BibitemOpen
  \bibfield  {author} {\bibinfo {author} {\bibfnamefont {S.~N.}\ \bibnamefont
  {Dorogovtsev}}, \bibinfo {author} {\bibfnamefont {J.~F.~F.}\ \bibnamefont
  {Mendes}}, \ and\ \bibinfo {author} {\bibfnamefont {A.~N.}\ \bibnamefont
  {Samukhin}},\ }\href {\doibase 10.1103/PhysRevLett.85.4633} {\bibfield
  {journal} {\bibinfo  {journal} {Phys. Rev. Lett.}\ }\textbf {\bibinfo
  {volume} {85}},\ \bibinfo {pages} {4633} (\bibinfo {year}
  {2000})}\BibitemShut {NoStop}%
\bibitem [{\citenamefont {Dorogovtsev}\ and\ \citenamefont
  {Mendes}(2001)}]{DorogovtsevMendes_2001_continuous}%
  \BibitemOpen
  \bibfield  {author} {\bibinfo {author} {\bibfnamefont {S.~N.}\ \bibnamefont
  {Dorogovtsev}}\ and\ \bibinfo {author} {\bibfnamefont {J.~F.~F.}\
  \bibnamefont {Mendes}},\ }\href {\doibase 10.1103/PhysRevE.63.056125}
  {\bibfield  {journal} {\bibinfo  {journal} {Phys. Rev. E}\ }\textbf {\bibinfo
  {volume} {63}},\ \bibinfo {pages} {056125} (\bibinfo {year}
  {2001})}\BibitemShut {NoStop}%
\bibitem [{\citenamefont {Dorogovtsev}\ and\ \citenamefont
  {Mendes}(2000)}]{DorogovtsevMendes_2000_aging}%
  \BibitemOpen
  \bibfield  {author} {\bibinfo {author} {\bibfnamefont {S.~N.}\ \bibnamefont
  {Dorogovtsev}}\ and\ \bibinfo {author} {\bibfnamefont {J.~F.~F.}\
  \bibnamefont {Mendes}},\ }\href {\doibase 10.1103/PhysRevE.62.1842}
  {\bibfield  {journal} {\bibinfo  {journal} {Phys. Rev. E}\ }\textbf {\bibinfo
  {volume} {62}},\ \bibinfo {pages} {1842} (\bibinfo {year}
  {2000})}\BibitemShut {NoStop}%
\bibitem [{\citenamefont {Bianconi}\ and\ \citenamefont
  {Barab\'asi}(2001)}]{BianconiBarabasi_2001}%
  \BibitemOpen
  \bibfield  {author} {\bibinfo {author} {\bibfnamefont {G.}~\bibnamefont
  {Bianconi}}\ and\ \bibinfo {author} {\bibfnamefont {A.-L.}\ \bibnamefont
  {Barab\'asi}},\ }\href {\doibase 10.1103/PhysRevLett.86.5632} {\bibfield
  {journal} {\bibinfo  {journal} {Phys. Rev. Lett.}\ }\textbf {\bibinfo
  {volume} {86}},\ \bibinfo {pages} {5632} (\bibinfo {year}
  {2001})}\BibitemShut {NoStop}%
\bibitem [{\citenamefont {Krapivsky}\ and\ \citenamefont
  {Redner}(2001)}]{KrapivskyRedner_2001}%
  \BibitemOpen
  \bibfield  {author} {\bibinfo {author} {\bibfnamefont {P.~L.}\ \bibnamefont
  {Krapivsky}}\ and\ \bibinfo {author} {\bibfnamefont {S.}~\bibnamefont
  {Redner}},\ }\href {\doibase 10.1103/PhysRevE.63.066123} {\bibfield
  {journal} {\bibinfo  {journal} {Phys. Rev. E}\ }\textbf {\bibinfo {volume}
  {63}},\ \bibinfo {pages} {066123} (\bibinfo {year} {2001})}\BibitemShut
  {NoStop}%
\bibitem [{\citenamefont {Krapivsky}\ \emph {et~al.}(2000)\citenamefont
  {Krapivsky}, \citenamefont {Redner},\ and\ \citenamefont
  {Leyvraz}}]{KrapivskyRednerLeyvraz_2000}%
  \BibitemOpen
  \bibfield  {author} {\bibinfo {author} {\bibfnamefont {P.~L.}\ \bibnamefont
  {Krapivsky}}, \bibinfo {author} {\bibfnamefont {S.}~\bibnamefont {Redner}}, \
  and\ \bibinfo {author} {\bibfnamefont {F.}~\bibnamefont {Leyvraz}},\ }\href
  {\doibase 10.1103/PhysRevLett.85.4629} {\bibfield  {journal} {\bibinfo
  {journal} {Phys. Rev. Lett.}\ }\textbf {\bibinfo {volume} {85}},\ \bibinfo
  {pages} {4629} (\bibinfo {year} {2000})}\BibitemShut {NoStop}%
\bibitem [{\citenamefont {Radicchi}\ \emph {et~al.}(2008)\citenamefont
  {Radicchi}, \citenamefont {Fortunato},\ and\ \citenamefont
  {Castellano}}]{RadicchiFortunatoCastellano2008}%
  \BibitemOpen
  \bibfield  {author} {\bibinfo {author} {\bibfnamefont {F.}~\bibnamefont
  {Radicchi}}, \bibinfo {author} {\bibfnamefont {S.}~\bibnamefont {Fortunato}},
  \ and\ \bibinfo {author} {\bibfnamefont {C.}~\bibnamefont {Castellano}},\
  }\href {\doibase 10.1073/pnas.0806977105} {\bibfield  {journal} {\bibinfo
  {journal} {Proc. Natl. Acad. Sci. USA}\ }\textbf {\bibinfo {volume} {105}},\
  \bibinfo {pages} {17268} (\bibinfo {year} {2008})}\BibitemShut {NoStop}%
\bibitem [{\citenamefont {Bollen}\ \emph {et~al.}(2009)\citenamefont {Bollen},
  \citenamefont {Van~de Sompel}, \citenamefont {Hagberg},\ and\ \citenamefont
  {Chute}}]{Bollen_etal_2009}%
  \BibitemOpen
  \bibfield  {author} {\bibinfo {author} {\bibfnamefont {J.}~\bibnamefont
  {Bollen}}, \bibinfo {author} {\bibfnamefont {H.}~\bibnamefont {Van~de
  Sompel}}, \bibinfo {author} {\bibfnamefont {A.}~\bibnamefont {Hagberg}}, \
  and\ \bibinfo {author} {\bibfnamefont {R.}~\bibnamefont {Chute}},\ }\href
  {\doibase 10.1371/journal.pone.0006022} {\bibfield  {journal} {\bibinfo
  {journal} {PLoS ONE}\ }\textbf {\bibinfo {volume} {4}},\ \bibinfo {pages}
  {e6022} (\bibinfo {year} {2009})}\BibitemShut {NoStop}%
\bibitem [{\citenamefont {Chen}\ \emph {et~al.}(2007)\citenamefont {Chen},
  \citenamefont {Xie}, \citenamefont {Maslov},\ and\ \citenamefont
  {Redner}}]{chen_etal_2006}%
  \BibitemOpen
  \bibfield  {author} {\bibinfo {author} {\bibfnamefont {P.}~\bibnamefont
  {Chen}}, \bibinfo {author} {\bibfnamefont {H.}~\bibnamefont {Xie}}, \bibinfo
  {author} {\bibfnamefont {S.}~\bibnamefont {Maslov}}, \ and\ \bibinfo {author}
  {\bibfnamefont {S.}~\bibnamefont {Redner}},\ }\href {\doibase
  10.1016/j.joi.2006.06.001} {\bibfield  {journal} {\bibinfo  {journal}
  {Journal of Informetrics}\ }\textbf {\bibinfo {volume} {1}},\ \bibinfo
  {pages} {8 } (\bibinfo {year} {2007})}\BibitemShut {NoStop}%
\bibitem [{\citenamefont {Walker}\ \emph {et~al.}(2007)\citenamefont {Walker},
  \citenamefont {Xie}, \citenamefont {Yan},\ and\ \citenamefont
  {Maslov}}]{walker_etal_2007}%
  \BibitemOpen
  \bibfield  {author} {\bibinfo {author} {\bibfnamefont {D.}~\bibnamefont
  {Walker}}, \bibinfo {author} {\bibfnamefont {H.}~\bibnamefont {Xie}},
  \bibinfo {author} {\bibfnamefont {K.-K.}\ \bibnamefont {Yan}}, \ and\
  \bibinfo {author} {\bibfnamefont {S.}~\bibnamefont {Maslov}},\ }\href
  {http://stacks.iop.org/1742-5468/2007/i=06/a=P06010} {\bibfield  {journal}
  {\bibinfo  {journal} {Journal of Statistical Mechanics: Theory and
  Experiment}\ }\textbf {\bibinfo {volume} {2007}},\ \bibinfo {pages} {P06010}
  (\bibinfo {year} {2007})}\BibitemShut {NoStop}%
\bibitem [{\citenamefont {Klosik}\ and\ \citenamefont
  {Bornholdt}(2013)}]{klosik_citation_2013}%
  \BibitemOpen
  \bibfield  {author} {\bibinfo {author} {\bibfnamefont {D.~F.}\ \bibnamefont
  {Klosik}}\ and\ \bibinfo {author} {\bibfnamefont {S.}~\bibnamefont
  {Bornholdt}},\ }\href {http://arxiv.org/abs/1301.7471} {\bibfield  {journal}
  {\bibinfo  {journal} {{arXiv:1301.7471}}\ } (\bibinfo {year}
  {2013})}\BibitemShut {NoStop}%
\bibitem [{\citenamefont {Hirsch}(2005)}]{Hirsch2005}%
  \BibitemOpen
  \bibfield  {author} {\bibinfo {author} {\bibfnamefont {J.~E.}\ \bibnamefont
  {Hirsch}},\ }\href {\doibase 10.1073/pnas.0507655102} {\bibfield  {journal}
  {\bibinfo  {journal} {Proc. Natl Acad. Sci. USA}\ }\textbf {\bibinfo {volume}
  {102}},\ \bibinfo {pages} {16569} (\bibinfo {year} {2005})}\BibitemShut
  {NoStop}%
\bibitem [{\citenamefont {Radicchi}\ \emph {et~al.}(2009)\citenamefont
  {Radicchi}, \citenamefont {Fortunato}, \citenamefont {Markines},\ and\
  \citenamefont {Vespignani}}]{Radicchi_et_al_2009}%
  \BibitemOpen
  \bibfield  {author} {\bibinfo {author} {\bibfnamefont {F.}~\bibnamefont
  {Radicchi}}, \bibinfo {author} {\bibfnamefont {S.}~\bibnamefont {Fortunato}},
  \bibinfo {author} {\bibfnamefont {B.}~\bibnamefont {Markines}}, \ and\
  \bibinfo {author} {\bibfnamefont {A.}~\bibnamefont {Vespignani}},\ }\href
  {\doibase 10.1103/PhysRevE.80.056103} {\bibfield  {journal} {\bibinfo
  {journal} {Phys. Rev. E}\ }\textbf {\bibinfo {volume} {80}},\ \bibinfo
  {pages} {056103} (\bibinfo {year} {2009})}\BibitemShut {NoStop}%
\bibitem [{\citenamefont {West}\ \emph {et~al.}(2010)\citenamefont {West},
  \citenamefont {Bergstrom},\ and\ \citenamefont {Bergstrom}}]{West_etal_2010}%
  \BibitemOpen
  \bibfield  {author} {\bibinfo {author} {\bibfnamefont {J.~D.}\ \bibnamefont
  {West}}, \bibinfo {author} {\bibfnamefont {T.~C.}\ \bibnamefont {Bergstrom}},
  \ and\ \bibinfo {author} {\bibfnamefont {C.~T.}\ \bibnamefont {Bergstrom}},\
  }\href {\doibase 10.5860/0710236} {\bibfield  {journal} {\bibinfo  {journal}
  {College \& Research Libraries}\ }\textbf {\bibinfo {volume} {71}},\ \bibinfo
  {pages} {236} (\bibinfo {year} {2010})}\BibitemShut {NoStop}%
\bibitem [{\citenamefont {Chen}\ and\ \citenamefont
  {Redner}(2010)}]{chen_redner_2010}%
  \BibitemOpen
  \bibfield  {author} {\bibinfo {author} {\bibfnamefont {P.}~\bibnamefont
  {Chen}}\ and\ \bibinfo {author} {\bibfnamefont {S.}~\bibnamefont {Redner}},\
  }\href {\doibase 10.1016/j.joi.2010.01.001} {\bibfield  {journal} {\bibinfo
  {journal} {Journal of Informetrics}\ }\textbf {\bibinfo {volume} {4}},\
  \bibinfo {pages} {278 } (\bibinfo {year} {2010})}\BibitemShut {NoStop}%
\bibitem [{\citenamefont {Wu}\ and\ \citenamefont
  {Holme}(2009)}]{WuHolme_2009}%
  \BibitemOpen
  \bibfield  {author} {\bibinfo {author} {\bibfnamefont {Z.-X.}\ \bibnamefont
  {Wu}}\ and\ \bibinfo {author} {\bibfnamefont {P.}~\bibnamefont {Holme}},\
  }\href {\doibase 10.1103/PhysRevE.80.037101} {\bibfield  {journal} {\bibinfo
  {journal} {Phys. Rev. E}\ }\textbf {\bibinfo {volume} {80}},\ \bibinfo
  {pages} {037101} (\bibinfo {year} {2009})}\BibitemShut {NoStop}%
\bibitem [{\citenamefont
  {Newman}(2001{\natexlab{a}})}]{newman_collaboration_I}%
  \BibitemOpen
  \bibfield  {author} {\bibinfo {author} {\bibfnamefont {M.~E.~J.}\
  \bibnamefont {Newman}},\ }\href {\doibase 10.1103/PhysRevE.64.016131}
  {\bibfield  {journal} {\bibinfo  {journal} {Physical Review E}\ }\textbf
  {\bibinfo {volume} {64}},\ \bibinfo {pages} {016131} (\bibinfo {year}
  {2001}{\natexlab{a}})}\BibitemShut {NoStop}%
\bibitem [{\citenamefont
  {Newman}(2001{\natexlab{b}})}]{newman_collaboration_II}%
  \BibitemOpen
  \bibfield  {author} {\bibinfo {author} {\bibfnamefont {M.~E.~J.}\
  \bibnamefont {Newman}},\ }\href {\doibase 10.1103/PhysRevE.64.016132}
  {\bibfield  {journal} {\bibinfo  {journal} {Physical Review E}\ }\textbf
  {\bibinfo {volume} {64}},\ \bibinfo {pages} {016132} (\bibinfo {year}
  {2001}{\natexlab{b}})}\BibitemShut {NoStop}%
\bibitem [{\citenamefont {Martin}\ \emph {et~al.}(2013)\citenamefont {Martin},
  \citenamefont {Ball}, \citenamefont {Karrer},\ and\ \citenamefont
  {Newman}}]{martin_coauthorship_2013}%
  \BibitemOpen
  \bibfield  {author} {\bibinfo {author} {\bibfnamefont {T.}~\bibnamefont
  {Martin}}, \bibinfo {author} {\bibfnamefont {B.}~\bibnamefont {Ball}},
  \bibinfo {author} {\bibfnamefont {B.}~\bibnamefont {Karrer}}, \ and\ \bibinfo
  {author} {\bibfnamefont {M.~E.~J.}\ \bibnamefont {Newman}},\ }\href {\doibase
  10.1103/PhysRevE.88.012814} {\bibfield  {journal} {\bibinfo  {journal}
  {Physical Review E}\ }\textbf {\bibinfo {volume} {88}},\ \bibinfo {pages}
  {012814} (\bibinfo {year} {2013})}\BibitemShut {NoStop}%
\bibitem [{\citenamefont {Krumov}\ \emph {et~al.}(2011)\citenamefont {Krumov},
  \citenamefont {Fretter}, \citenamefont {M\"uller-Hannemann}, \citenamefont
  {Weihe},\ and\ \citenamefont {H\"utt}}]{krumov_motifs_2011}%
  \BibitemOpen
  \bibfield  {author} {\bibinfo {author} {\bibfnamefont {L.}~\bibnamefont
  {Krumov}}, \bibinfo {author} {\bibfnamefont {C.}~\bibnamefont {Fretter}},
  \bibinfo {author} {\bibfnamefont {M.}~\bibnamefont {M\"uller-Hannemann}},
  \bibinfo {author} {\bibfnamefont {K.}~\bibnamefont {Weihe}}, \ and\ \bibinfo
  {author} {\bibfnamefont {M.-T.}\ \bibnamefont {H\"utt}},\ }\href {\doibase
  10.1140/epjb/e2011-10746-5} {\bibfield  {journal} {\bibinfo  {journal} {The
  European Physical Journal B}\ }\textbf {\bibinfo {volume} {84}},\ \bibinfo
  {pages} {535} (\bibinfo {year} {2011})}\BibitemShut {NoStop}%
\bibitem [{\citenamefont {Milo}\ \emph {et~al.}(2002)\citenamefont {Milo},
  \citenamefont {Shen-Orr}, \citenamefont {Itzkovitz}, \citenamefont {Kashtan},
  \citenamefont {Chklovskii},\ and\ \citenamefont {Alon}}]{milo_motifs_2002}%
  \BibitemOpen
  \bibfield  {author} {\bibinfo {author} {\bibfnamefont {R.}~\bibnamefont
  {Milo}}, \bibinfo {author} {\bibfnamefont {S.}~\bibnamefont {Shen-Orr}},
  \bibinfo {author} {\bibfnamefont {S.}~\bibnamefont {Itzkovitz}}, \bibinfo
  {author} {\bibfnamefont {N.}~\bibnamefont {Kashtan}}, \bibinfo {author}
  {\bibfnamefont {D.}~\bibnamefont {Chklovskii}}, \ and\ \bibinfo {author}
  {\bibfnamefont {U.}~\bibnamefont {Alon}},\ }\href {\doibase
  10.1126/science.298.5594.824} {\bibfield  {journal} {\bibinfo  {journal}
  {Science}\ }\textbf {\bibinfo {volume} {298}},\ \bibinfo {pages} {824}
  (\bibinfo {year} {2002})}\BibitemShut {NoStop}%
\bibitem [{\citenamefont {Milo}\ \emph {et~al.}(2004)\citenamefont {Milo},
  \citenamefont {Itzkovitz}, \citenamefont {Kashtan}, \citenamefont {Levitt},
  \citenamefont {Shen-Orr}, \citenamefont {Ayzenshtat}, \citenamefont
  {Sheffer},\ and\ \citenamefont {Alon}}]{milo_superfamilies_2004}%
  \BibitemOpen
  \bibfield  {author} {\bibinfo {author} {\bibfnamefont {R.}~\bibnamefont
  {Milo}}, \bibinfo {author} {\bibfnamefont {S.}~\bibnamefont {Itzkovitz}},
  \bibinfo {author} {\bibfnamefont {N.}~\bibnamefont {Kashtan}}, \bibinfo
  {author} {\bibfnamefont {R.}~\bibnamefont {Levitt}}, \bibinfo {author}
  {\bibfnamefont {S.}~\bibnamefont {Shen-Orr}}, \bibinfo {author}
  {\bibfnamefont {I.}~\bibnamefont {Ayzenshtat}}, \bibinfo {author}
  {\bibfnamefont {M.}~\bibnamefont {Sheffer}}, \ and\ \bibinfo {author}
  {\bibfnamefont {U.}~\bibnamefont {Alon}},\ }\href {\doibase
  10.1126/science.1089167} {\bibfield  {journal} {\bibinfo  {journal}
  {Science}\ }\textbf {\bibinfo {volume} {303}},\ \bibinfo {pages} {1538}
  (\bibinfo {year} {2004})}\BibitemShut {NoStop}%
\bibitem [{\citenamefont {Fretter}\ \emph {et~al.}(2012)\citenamefont
  {Fretter}, \citenamefont {M{\"u}ller-Hannemann},\ and\ \citenamefont
  {H{\"u}tt}}]{fretter_subgraph_2012}%
  \BibitemOpen
  \bibfield  {author} {\bibinfo {author} {\bibfnamefont {C.}~\bibnamefont
  {Fretter}}, \bibinfo {author} {\bibfnamefont {M.}~\bibnamefont
  {M{\"u}ller-Hannemann}}, \ and\ \bibinfo {author} {\bibfnamefont {M.-T.}\
  \bibnamefont {H{\"u}tt}},\ }\href {\doibase 10.1103/PhysRevE.85.056119}
  {\bibfield  {journal} {\bibinfo  {journal} {Physical Review E}\ }\textbf
  {\bibinfo {volume} {85}},\ \bibinfo {pages} {056119} (\bibinfo {year}
  {2012})}\BibitemShut {NoStop}%
\bibitem [{\citenamefont {Beber}\ \emph {et~al.}(2012)\citenamefont {Beber},
  \citenamefont {Fretter}, \citenamefont {Jain}, \citenamefont {Sonnenschein},
  \citenamefont {M{\"u}ller-Hannemann},\ and\ \citenamefont
  {H{\"u}tt}}]{beber_artefacts_2012}%
  \BibitemOpen
  \bibfield  {author} {\bibinfo {author} {\bibfnamefont {M.~E.}\ \bibnamefont
  {Beber}}, \bibinfo {author} {\bibfnamefont {C.}~\bibnamefont {Fretter}},
  \bibinfo {author} {\bibfnamefont {S.}~\bibnamefont {Jain}}, \bibinfo {author}
  {\bibfnamefont {N.}~\bibnamefont {Sonnenschein}}, \bibinfo {author}
  {\bibfnamefont {M.}~\bibnamefont {M{\"u}ller-Hannemann}}, \ and\ \bibinfo
  {author} {\bibfnamefont {M.-T.}\ \bibnamefont {H{\"u}tt}},\ }\href {\doibase
  10.1098/rsif.2012.0490} {\bibfield  {journal} {\bibinfo  {journal} {Journal
  of The Royal Society Interface}\ }\textbf {\bibinfo {volume} {9}},\ \bibinfo
  {pages} {3426} (\bibinfo {year} {2012})}\BibitemShut {NoStop}%
\bibitem [{\citenamefont {Schulz}\ \emph {et~al.}(2014)\citenamefont {Schulz},
  \citenamefont {Mazloumian}, \citenamefont {Petersen}, \citenamefont
  {Penner},\ and\ \citenamefont {Helbing}}]{schulz_exploiting_2014}%
  \BibitemOpen
  \bibfield  {author} {\bibinfo {author} {\bibfnamefont {C.}~\bibnamefont
  {Schulz}}, \bibinfo {author} {\bibfnamefont {A.}~\bibnamefont {Mazloumian}},
  \bibinfo {author} {\bibfnamefont {A.~M.}\ \bibnamefont {Petersen}}, \bibinfo
  {author} {\bibfnamefont {O.}~\bibnamefont {Penner}}, \ and\ \bibinfo {author}
  {\bibfnamefont {D.}~\bibnamefont {Helbing}},\ }\href
  {http://arxiv.org/abs/1401.6157} {\bibfield  {journal} {\bibinfo  {journal}
  {{arXiv:1401.6157}}\ } (\bibinfo {year} {2014})}\BibitemShut {NoStop}%
\bibitem [{\citenamefont {Christiano~Silva}\ and\ \citenamefont
  {Raphael~Amancio}(2013)}]{silva_network-based_2013}%
  \BibitemOpen
  \bibfield  {author} {\bibinfo {author} {\bibfnamefont {T.}~\bibnamefont
  {Christiano~Silva}}\ and\ \bibinfo {author} {\bibfnamefont {D.}~\bibnamefont
  {Raphael~Amancio}},\ }\href {\doibase http://dx.doi.org/10.1063/1.4794795}
  {\bibfield  {journal} {\bibinfo  {journal} {Chaos}\ }\textbf {\bibinfo
  {volume} {23}},\ \bibinfo {eid} {013139} (\bibinfo {year}
  {2013})}\BibitemShut {NoStop}%
\bibitem [{\citenamefont {Fegley}\ and\ \citenamefont
  {Torvik}(2013)}]{fegleyTorvik_2013}%
  \BibitemOpen
  \bibfield  {author} {\bibinfo {author} {\bibfnamefont {B.~D.}\ \bibnamefont
  {Fegley}}\ and\ \bibinfo {author} {\bibfnamefont {V.~I.}\ \bibnamefont
  {Torvik}},\ }\href {\doibase 10.1371/journal.pone.0070299} {\bibfield
  {journal} {\bibinfo  {journal} {PLoS ONE}\ }\textbf {\bibinfo {volume} {8}},\
  \bibinfo {pages} {e70299} (\bibinfo {year} {2013})}\BibitemShut {NoStop}%
\bibitem [{\citenamefont {Milojevi{\c c}}(2013)}]{Milojevic2013}%
  \BibitemOpen
  \bibfield  {author} {\bibinfo {author} {\bibfnamefont {S.}~\bibnamefont
  {Milojevi{\c c}}},\ }\href {\doibase
  http://dx.doi.org/10.1016/j.joi.2013.06.006} {\bibfield  {journal} {\bibinfo
  {journal} {Journal of Informetrics}\ }\textbf {\bibinfo {volume} {7}},\
  \bibinfo {pages} {767 } (\bibinfo {year} {2013})}\BibitemShut {NoStop}%
\bibitem [{Note1()}]{Note1}%
  \BibitemOpen
  \bibinfo {note} {See \protect \url
  {https://publish.aps.org/datasets}.}\BibitemShut {Stop}%
\bibitem [{\citenamefont {Wernicke}(2006)}]{wernicke}%
  \BibitemOpen
  \bibfield  {author} {\bibinfo {author} {\bibfnamefont {S.}~\bibnamefont
  {Wernicke}},\ }\href {\doibase 10.1109/TCBB.2006.51} {\bibfield  {journal}
  {\bibinfo  {journal} {{IEEE/ACM} Transactions on Computational Biology and
  Bioinformatics}\ }\textbf {\bibinfo {volume} {3}},\ \bibinfo {pages} {347}
  (\bibinfo {year} {2006})}\BibitemShut {NoStop}%
\bibitem [{\citenamefont {Loss}\ and\ \citenamefont
  {DiVincenzo}(1998)}]{PhysRevA.57.120}%
  \BibitemOpen
  \bibfield  {author} {\bibinfo {author} {\bibfnamefont {D.}~\bibnamefont
  {Loss}}\ and\ \bibinfo {author} {\bibfnamefont {D.~P.}\ \bibnamefont
  {DiVincenzo}},\ }\href {\doibase 10.1103/PhysRevA.57.120} {\bibfield
  {journal} {\bibinfo  {journal} {Phys. Rev. A}\ }\textbf {\bibinfo {volume}
  {57}},\ \bibinfo {pages} {120} (\bibinfo {year} {1998})}\BibitemShut
  {NoStop}%
\bibitem [{\citenamefont {Feist}\ \emph {et~al.}(2009)\citenamefont {Feist},
  \citenamefont {Herrg{\aa}rd}, \citenamefont {Thiele}, \citenamefont {Reed},\
  and\ \citenamefont {Palsson}}]{Feist:2009ik}%
  \BibitemOpen
  \bibfield  {author} {\bibinfo {author} {\bibfnamefont {A.~M.}\ \bibnamefont
  {Feist}}, \bibinfo {author} {\bibfnamefont {M.~J.}\ \bibnamefont
  {Herrg{\aa}rd}}, \bibinfo {author} {\bibfnamefont {I.}~\bibnamefont
  {Thiele}}, \bibinfo {author} {\bibfnamefont {J.~L.}\ \bibnamefont {Reed}}, \
  and\ \bibinfo {author} {\bibfnamefont {B.~{\O}.}\ \bibnamefont {Palsson}},\
  }\href {\doibase doi:10.1038/nrmicro1949} {\bibfield  {journal} {\bibinfo
  {journal} {Nature Reviews Microbiology}\ }\textbf {\bibinfo {volume} {7}},\
  \bibinfo {pages} {129} (\bibinfo {year} {2009})}\BibitemShut {NoStop}%
\bibitem [{\citenamefont {Schellenberger}\ \emph {et~al.}(2010)\citenamefont
  {Schellenberger}, \citenamefont {Park}, \citenamefont {Conrad},\ and\
  \citenamefont {Palsson}}]{Schellenberger:2010kb}%
  \BibitemOpen
  \bibfield  {author} {\bibinfo {author} {\bibfnamefont {J.}~\bibnamefont
  {Schellenberger}}, \bibinfo {author} {\bibfnamefont {J.~O.}\ \bibnamefont
  {Park}}, \bibinfo {author} {\bibfnamefont {T.~M.}\ \bibnamefont {Conrad}}, \
  and\ \bibinfo {author} {\bibfnamefont {B.~{\O}.}\ \bibnamefont {Palsson}},\
  }\href {\doibase 10.1186/1471-2105-11-213} {\bibfield  {journal} {\bibinfo
  {journal} {BMC Bioinformatics}\ }\textbf {\bibinfo {volume} {11}},\ \bibinfo
  {pages} {213} (\bibinfo {year} {2010})}\BibitemShut {NoStop}%
\bibitem [{\citenamefont {Marr}\ \emph {et~al.}(2008)\citenamefont {Marr},
  \citenamefont {Geertz}, \citenamefont {H{\"u}tt},\ and\ \citenamefont
  {Muskhelishvili}}]{Marr:2008p4284}%
  \BibitemOpen
  \bibfield  {author} {\bibinfo {author} {\bibfnamefont {C.}~\bibnamefont
  {Marr}}, \bibinfo {author} {\bibfnamefont {M.}~\bibnamefont {Geertz}},
  \bibinfo {author} {\bibfnamefont {M.-T.}\ \bibnamefont {H{\"u}tt}}, \ and\
  \bibinfo {author} {\bibfnamefont {G.}~\bibnamefont {Muskhelishvili}},\ }\href
  {\doibase 10.1186/1752-0509-2-18} {\bibfield  {journal} {\bibinfo  {journal}
  {BMC Syst Biol}\ }\textbf {\bibinfo {volume} {2}},\ \bibinfo {pages} {18}
  (\bibinfo {year} {2008})}\BibitemShut {NoStop}%
\bibitem [{\citenamefont {Blumenthal}(2004)}]{blumenthal2004}%
  \BibitemOpen
  \bibfield  {author} {\bibinfo {author} {\bibfnamefont {T.}~\bibnamefont
  {Blumenthal}},\ }\href {\doibase 10.1093/bfgp/3.3.199} {\bibfield  {journal}
  {\bibinfo  {journal} {Briefings in Functional Genomics and Proteomics}\
  }\textbf {\bibinfo {volume} {3}},\ \bibinfo {pages} {199} (\bibinfo {year}
  {2004})}\BibitemShut {NoStop}%
\bibitem [{\citenamefont {Hagmann}\ \emph {et~al.}(2008)\citenamefont
  {Hagmann}, \citenamefont {Cammoun}, \citenamefont {Gigandet}, \citenamefont
  {Meuli}, \citenamefont {Honey}, \citenamefont {Wedeen},\ and\ \citenamefont
  {Sporns}}]{Hagmann2008}%
  \BibitemOpen
  \bibfield  {author} {\bibinfo {author} {\bibfnamefont {P.}~\bibnamefont
  {Hagmann}}, \bibinfo {author} {\bibfnamefont {L.}~\bibnamefont {Cammoun}},
  \bibinfo {author} {\bibfnamefont {X.}~\bibnamefont {Gigandet}}, \bibinfo
  {author} {\bibfnamefont {R.}~\bibnamefont {Meuli}}, \bibinfo {author}
  {\bibfnamefont {C.~J.}\ \bibnamefont {Honey}}, \bibinfo {author}
  {\bibfnamefont {V.~J.}\ \bibnamefont {Wedeen}}, \ and\ \bibinfo {author}
  {\bibfnamefont {O.}~\bibnamefont {Sporns}},\ }\href {\doibase 07-PLBI-RA-4028
  [pii] 10.1371/journal.pbio.0060159} {\bibfield  {journal} {\bibinfo
  {journal} {PLoS Biol}\ }\textbf {\bibinfo {volume} {6}},\ \bibinfo {pages}
  {e159} (\bibinfo {year} {2008})}\BibitemShut {NoStop}%
\end{thebibliography}%

\end{document}